\documentclass{INTERSPEECH2023}

\interspeechcameraready 

\usepackage{preamble}

\title{Investigating Reproducibility at Interspeech Conferences:\\A Longitudinal and Comparative Perspective}

\name{Mohammad Arvan,$^1$ A. Seza Doğruöz,$^2$ Natalie Parde$^1$}

\address{
  $^1$University of Illinois Chicago, USA\\
  $^2$Ghent University, Belgium}
\email{marvan3@uic.edu, as.dogruoz@ugent.be, parde@uic.edu}

\begin{document}

\maketitle

\begin{abstract}
  Reproducibility is a key aspect for scientific advancement across disciplines, and reducing barriers for open science is a focus area for the theme of Interspeech 2023. Availability of source code is one of the indicators that facilitates reproducibility. However, less is known about the rates of reproducibility at Interspeech conferences in comparison to other conferences in the field. In order to fill this gap, we have surveyed 27,717 papers at seven conferences across speech and language processing disciplines.  We find that despite having a close number of accepted papers to the other conferences, Interspeech has up to 40\% less source code availability. In addition to reporting the difficulties we have encountered during our research, we also provide recommendations and possible directions to increase reproducibility for further studies.

  % (XX) based on the following criteria: XX, XX. 
  % In addition, we have also made a comparison with the rate of reproducibility for other conferences (XXX). As a result of this comparison, our findings suggest XX rate for the reproducibility of Interspeech publications and XX rates for XX conferences. 
\end{abstract}

%\ma{You can add comments by using $sd$ command}
%\sd{is this me? :)} 

\section{Introduction}

% Inclusive Spoken Language Science and Technology
% report barriers that could prevent other researchers adopting a technique, or users from benefitting
\let\thefootnote\relax\footnotetext{Published at Interspeech'23, DOI: 10.21437/Interspeech.2023-2252 }

The United Nations Educational, Scientific and Cultural Organization (UNESCO) has recently released its recommendations on open science \cite{unesco_2021_5834767}. These recommendations are based on the consideration that science's efficiency, effectiveness, and impact can be improved by making scientific knowledge, data, and information openly available, accessible, and reusable. Considering the theme of this year's Interspeech is to address barriers that could prevent other researchers from adopting a technique, or users from benefitting, we find this an excellent opportunity to discuss the importance of open science within the Interspeech community and highlight the ways it can mitigate such barriers. Part of these recommendations is educating researchers on ways to align their research with open science practices.

Fortunately, unlike many other disciplines, access to the scientific papers published at Interspeech is not restricted behind a paywall. Nevertheless, there may be other concerns, such as the reproducibility of the results reported in these papers. Since published papers may not include all the details required to reproduce the results, the availability and quality of the software and the data used in the experiments are also important factors to consider. The growing concerns regarding this matter, often referred to as the ``reproducibility crisis'' \cite{baker20161} has led to several action items to address this issue. 

To address this concern, the standardization of research- and publication-oriented best practices and the pursuit of reproducible research has been increasingly central within the machine learning and speech and language processing communities. Several top-tier conferences have promoted \textit{reproducibility challenges} as part of workshops \cite{ml_reproducibility_challenge} or even in initiatives in which authors can earn badges for their papers \cite{naacl_reproducibility_track}. Others have produced comprehensive \textit{reproducibility studies} to draw attention to the matter and suggest promising directions for improving conditions \cite{belz-etal-2022-quantified}. Collectively, these efforts have increased awareness of the importance of research reproducibility. Evidence suggests that they have also resulted in real gains in the rate at which authors release their source code or data \cite{DBLP:conf/emnlp/ArvanPP22}.

% Research in the speech and language processing community has progressed at a rapid pace in recent years, resulting in many exciting innovations and oftentimes producing performance measures that climb at dizzying rates.  However, with this increased pace of research comes increased difficulty in keeping up with an onslaught of intriguing publications, and moreover, in externally validating their findings.  Although it is one thing to report the state of the art, it is another thing entirely to provide sufficient material for justifying those claims.

Thus far, the extent to which these efforts have carried over to the speech processing community seems minimal. These efforts have been limited to asking authors to fill out a reproducibility checklist during the submission process. Additionally, this year's theme suggests that authors report performance metric distributions in addition to averages. In this work, we set out to systematically understand and report upon the state of reproducibility at Interspeech, the world's premier conference on speech processing, compared to peer conferences in the machine learning and natural language processing communities. 

%Our hypothesis is that the speech processing community suffers from large-scale reproducibility concerns that may be holding back the field from maximizing its potential.  We seek to prove (or disprove!) this hypothesis by studying trends in source code availability for Interspeech publications alongside those from its peer conferences in machine learning and NLP, as well as conducting case studies in which we attempt to reproduce a small selection of these papers ourselves.

Our primary contribution is to reveal current trends and promote recommendations for reducing identified barriers in the future, fostering more open science in the Interspeech community.  Although it is beyond the scope of this paper to comprehensively examine all facets of reproducibility within the community, we argue that our study captures a reasonable portrait of Interspeech reproducibility as a whole, as evidenced by relevant findings in comparable studies in machine learning and NLP.

\section{Related Work}

Although research towards building and sustaining reproducible and open science is gradually becoming more standardized, work to date on this topic has been approached from a wide variety of perspectives.  Because of this and also because of the novelty of this topic as research in its own right, different terms and definitions are often used in the context of reproducibility interchangeably. To avoid confusion and misunderstanding, we begin this section by briefly defining the terms used in our own work; namely, \textit{reproducibility}, \textit{replicability}, and \textit{open science}.  We follow these definitions with a brief summary of the evolution of work towards reproducibility and open science to date, framed within the broader AI community. We adapt the definitions originally proposed by The International Vocabulary of Metrology (VIM) \cite{jcgm2012jcgm}, highlighted by Belz et. al \cite{belz-etal-2022-quantified}. 

\textbf{Reproducibility}  Within the VIM framework, \textit{reproducibility} is defined as measurement precision under a set of conditions. Using precision enables the quantification of reproducibility by standard statistical measures such as the coefficient of variation (CV), computed as the ratio between the standard deviation $\sigma$ and mean $\mu$ of measured scores: $\frac{\sigma}{\mu}$. Specifically, when computing these statistical measures the conditions are all the variables affecting an observed result, including but not limited to source code, runtime environment, data, model, and hardware. 

\textbf{Replicability} Following the definition of reproducibility, we define \textit{replicability} (or \textit{repeatability}) as reproducibility under the same conditions. 

Although reproducibility may be framed as both a binary (yes/no) variable or a numeric score, the latter quantification of a reproduction attempt (most often using CV) may provide more information and be less subjective.  In a recent study \cite{DBLP:conf/eacl/BelzASR21}, only 14.03\% of 513 reproduction score pairs were found to be the same, highlighting the importance of more fine-grained quantitative measures in understanding the extent of reproducibility \cite{DBLP:journals/jmlr/PineauVSLBdFL21,whitaker_2017}.

% \textbf{Open Science.}
% \sd{Can we add the open science paper by Unesco and the definitions there as well? }

UNESCO \cite{unesco_2021_5834767} defines open science as ``an inclusive construct that combines various movements and practices that aim to make multilingual scientific knowledge openly available, accessible, and reusable for everyone.'' Many venues have encouraged researchers to share more of their work over the last few years \cite{nature_reporducibility,aaai_reporducibility,acm_reporducibility,acl_reporducibility,naacl_reproducibility_track}. Proposed checklists such as the ML Reproducibility Checklist \cite{ml_reproducibility_checklist} and the ML Completeness Checklist \cite{ml_completeness_checklist} ask authors to provide specification of dependencies, training and evaluation code, pre-trained models, and proper documentation on how to run the provided code. However, recent research suggests that this information may not be enough to support adequate reproducibility \cite{arvan-etal-2022-reproducibility,DBLP:conf/emnlp/ArvanPP22}.

One of the early assumptions with regard to reproducibility was that simply sharing source code, data, and hyperparameters would suffice to achieve reproducibility. More recent studies \cite{DBLP:conf/emnlp/ArvanPP22,DBLP:journals/corr/abs-2305-01633,arvan-etal-2022-reproducibility} have shown that in practice, however, this is not the case. In fact, even changing the random seed could have drastic impact on the final results \cite{DBLP:journals/corr/abs-2002-06305}. Controlling the conditions affecting results (and understanding what those conditions are) is extremely important to achieve reproducibility. Even with open source code, data, and hyperparameters, changes to other more innocuous or hidden experimental conditions may greatly influence reproducibility. For example, one of the factors missing from the original definition of reproducibility is the runtime environment. Listing the dependencies and the versions of the dependencies used when reporting results may foster greater reproducibility, but it also may not be sufficient for achieving reproducible results since packages may be updated or even become deprecated over time. Due to the nondeterministic nature of deep learning models, even the hardware used for a study can cause variation in the results. Fixing the random seed may fall short of addressing this concern as well as it is only guaranteed that it will generate the same sequence of numbers when used in the \textit{same} environment.

Ultimately, capturing all conditions involved in scientific experiments is an ever-evolving and challenging task. Currently, the best solution is to provide a self-contained docker container or a virtual machine image that can be used to reproduce the results. While this concept has not been fully adopted by the machine learning or NLP communities, it is considered as the best practice in other computing research areas such as systems and software engineering \cite{10.1145/3492349,DBLP:conf/asplos/2022}. While self-containment might not seem necessary at first glance, it prolongs the lifetime of the artifacts by reducing the chance of reliance on unavailable dependencies. 

% \textbf{Research Artifacts.} All of the materials released by authors to achieve reproducibility.
% \textbf{Self-Contained.} A container or runtime environment that includes all files and libraries required to run the specified source code, without any need to access the internet or any other external resources.

% According to Docker website \footnote{\url{https://www.docker.com/resources/what-container/}} itself, docker container is a lightweight, standalone, executable package of software that includes everything needed to run an application: code, runtime, system tools, system libraries and settings.

\section{Methodology}

Considering the necessity of access to research artifacts to most productively reproduce the results of computing research, including that within the speech processing community, we set out to study the availability of research artifacts for papers published at Interspeech. The papers published at Interspeech are available through the \url{isca-speech.org} website. This portal provides basic information regarding speech processing papers listed in the conference proceedings, as well as the papers themselves. At the time of writing, the website provides information regarding 35,050 papers published at 344 conferences. We limit our analysis to papers published at Interspeech from 2017 to 2022.  Given the fast pace of research in speech processing specifically and computing more broadly, we selected a five-year window to balance our competing interests in studying reproducibility over time while recognizing that the Interspeech community even just five years ago was in some ways very different from how it is today. We believe that 2019, the year that the NeurIPS 2019 Reproducibility Program started, was a turning point regarding the availability of research artifacts. Given our five-year window, we were able to include data that highlighted trends both prior to and following 2019.

% \begin{figure}
%     \centering
%     \includegraphics[width=0.9\columnwidth]{figures/Interspeech Methods.drawio.png}
%     \caption{Process for studying research availability across included conferences.} 
%     % \np{Added this figure since we had some room; feel free to edit or delete if we run out of space.}

%     \label{fig:study_process}
% \end{figure}

While the ISCA portal is useful for researchers who are searching for papers, it does not provide any information about the availability of research artifacts; hence, retrieving this information originally required us to manually search each paper. To expedite and streamline this process, we wrote a Python script to download the PDF files.  Then, we used the PyPDF2\footnote{\url{https://pypi.org/project/PyPDF2/}} library to extract the text from the PDFs. Following this, we performed a simple keyword search (``github.com'') to determine whether the paper contained any information regarding released software artifacts. GitHub is by and large the most popular website for hosting open source code, making its presence in a paper a reasonable first clue towards source code availability. The references section of the paper was excluded from the search to reduce false positives. 

To provide a point of comparison, we also collected the same information for major NLP and machine learning conferences (considered peer conferences to Interspeech in adjacent research fields). Unlike \url{isca-speech.org}, the \url{aclanthology.org} portal (which hosts the papers from all top-tier natural language processing conferences) contains ``code'' and ``data'' fields that indicate whether the paper contains a link to the source code and the data. We use these fields to determine the availability of research artifacts for these conferences. We limit our NLP search to papers published at ACL, EMNLP, NAACL, LREC, and COLING. For our machine learning search, we attempted to follow a similiar process for papers published at NeurIPS. This conference utilizes the \url{openreview.net} website to host their proceedings. The website provides an API to download accepted papers published at conferences. NeurIPS also provides a paper portal\footnote{\url{papers.neurips.cc}} that provides a similar experience to the ACL Anthology. 
% We summarize our study process in Figure \ref{fig:study_process}.

% The papers published at NeurIPS, ICLR, and ICML are available through the \url{openreview.net} website. Openreview provides an api download accepted papers published at conferences. 

% Paper selection is based on the following criteria:
% keyword search for ``multilinguality'' in search engines.

% Reproducibility assessment is based on the following criteria:

\section{Results}

\begin{figure}
  \centering
  \includegraphics[width=\columnwidth]{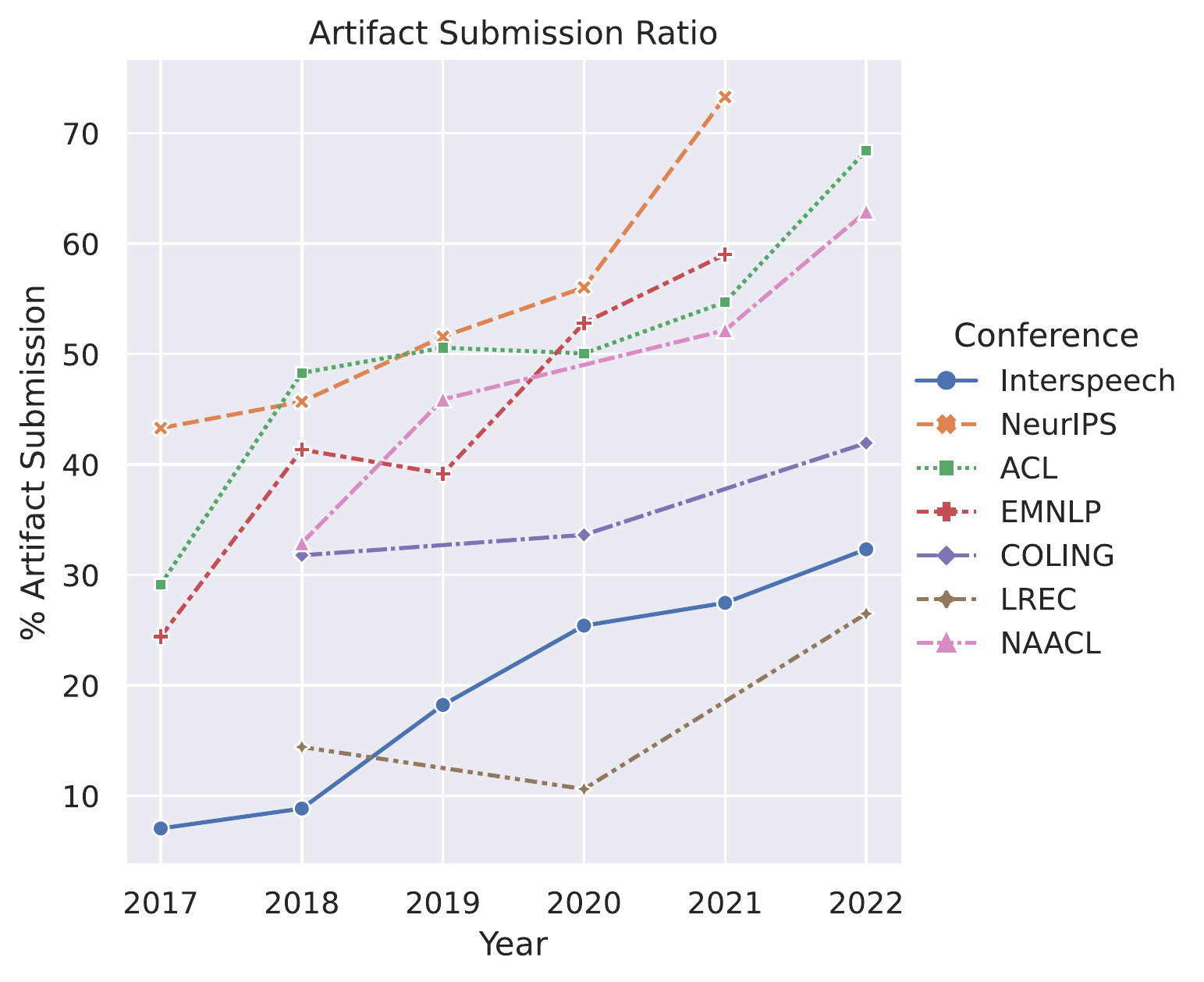}
  \caption{Percentage of papers with research artifacts at the selected conferences over the years.}
  \label{fig:code_inclusion}
\end{figure}

We present our primary results in Figure~\ref{fig:code_inclusion}. The figure shows the percentage of papers with research artifacts for each conference. We observe that the percentage of papers with research artifacts is higher in NeurIPS and nearly all of NLP conferences than in Interspeech. On a positive note, all of the conferences have an upward trend for research artifact submission. In the case of NeurIPS, ACL, EMNLP, and NAACL, the percentage has surpassed 50\%. The gap between research artifact availability in NeurIPS and Interspeech is nearly 40\%. In fact, except for LREC, all other conferences have a higher percentage of papers with research artifacts than Interspeech.

\begin{figure}
  \centering
  \includegraphics[width=\columnwidth]{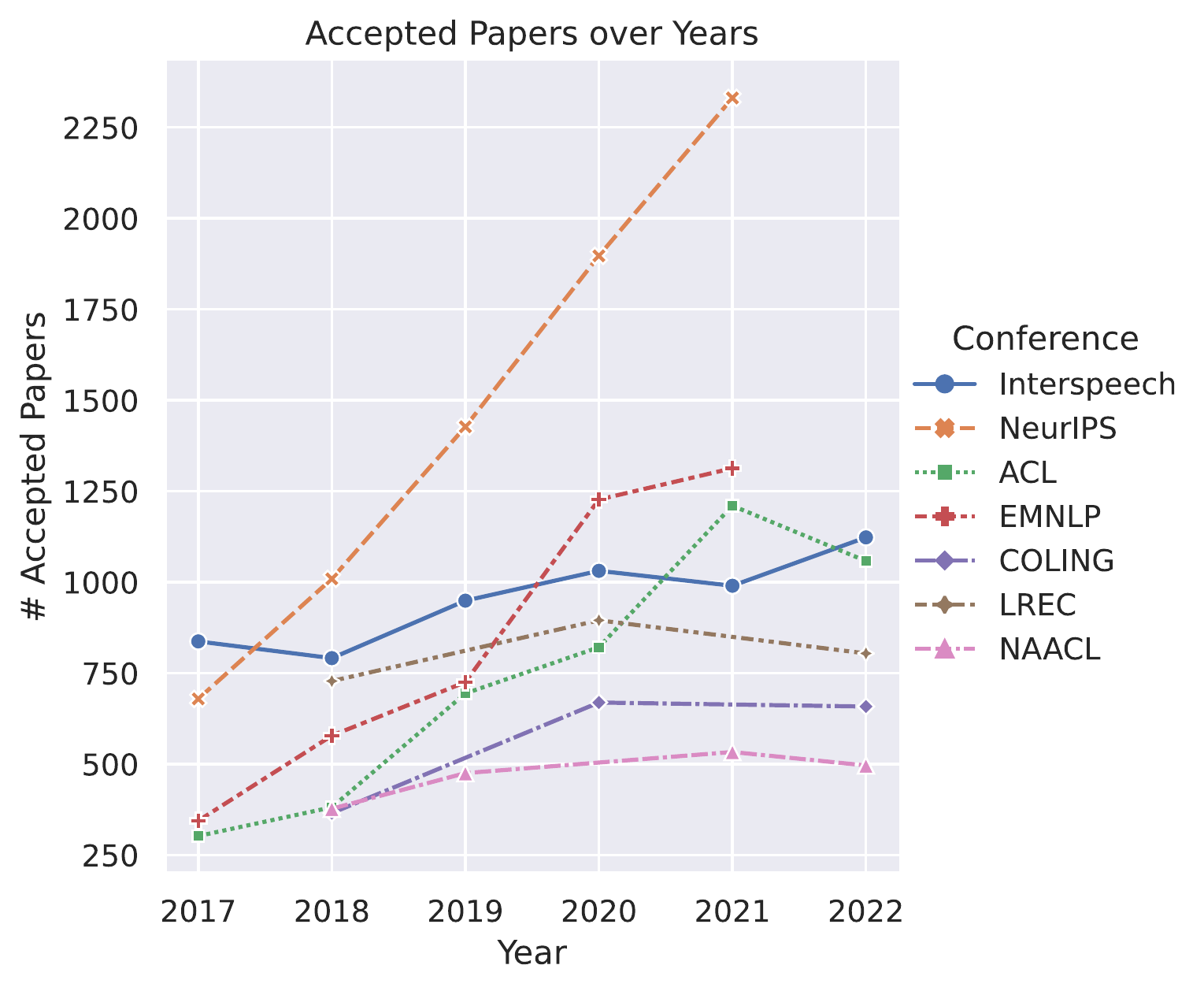}
  \caption{Total number of papers accepted at the selected conferences over the years.}
  \label{fig:total_papers}
\end{figure}

We demonstrate the total number of papers accepted at each conference in Figure~\ref{fig:total_papers}. Interestingly, in addition to being the front-runner in terms of its percentage of papers with research artifacts available, NeurIPS also has the greatest overall volume of papers accepted. The rest of the conferences are relatively close in terms of the total number of papers that they accepted.

The data presented was collected by scraping different portals. The issue with this approach is that if the structure or the design of the page changes, the entire collection process must be updated. Additionally, APIs often come with a limit rate to avoid overloading the backend servers.  Without having access to an API, we had to estimate our own limits to avoid overloading servers. While \url{openreview.net} comes with an API, there does not seem to be a way of differentiating between papers that have been accepted and those that have been rejected. Hence, we had to manually check the status of each paper. This process was time-consuming and error-prone. Had we been able to access that information through an API, we could have collected more data regarding machine learning conferences as well (e.g., by including ICML or ICLR), increasing the comprehensiveness of our analysis.

\section{Discussion}

\subsection{Where are we now?}

Based on our results, we can conclude that the percentage of papers with research artifacts in Interspeech is lower than that observed in other conferences. This is a concerning issue, as it may suggest a lack of reproducibility in research published at Interspeech. We believe this could create unnecessary barriers for future researchers to adapt and build upon the work presented at this conference. This problem is not easily solved, as it requires a change in established research practices and acceptance from the community. One may argue that it is harder to share research artifacts in the field of speech processing than other fields, due to issues such as complexity, artifact size, or even privacy concerns. However, we believe that the community should strive to make research artifacts available and that other communities have dealt with many parallel concerns; their solutions may offer guidance on how to address these concerns in the Interspeech community. We hope that this work will raise awareness about the importance of reproducibility in Interspeech. 

The \url{isca-speech.org} portal provides a useful resource for researchers to find papers. However, it does not provide any information about the availability of research artifacts. We believe that this portal should be updated to include this information. This would make it easier for researchers to find papers with research artifacts. 

% We refrain from making any recommendations on how to improve the situation, as earlier work has already provided some suggestions \cite{DBLP:conf/emnlp/ArvanPP22}.\np{Not sure if we want to refrain from making \textit{any} recommendations?}
% \ma{A.S. Dogruoz: Excellent but we need to suggest this a bit more specifically for the speech community}

\subsection{What is the next step?}

% Ideally, researchers should be able to reproduce the results of a research paper with a minimum amount of effort. When the conditions are clearly defined, it should be easy to assess whether the results are reproducible if one or more of the conditions are not maintained (in comparison to the original research paper). 
By identifying where we should be, we can provide a future direction. In an ideal scenario, future researchers would have access to fully functional self-contained artifacts alongside the report published by the original authors. This availability would facilitate further scrutiny and review. Furthermore, the research artifacts would contain details that may have been omitted in the paper itself, providing a comparison point in the event that other researchers chose to re-implement and reuse such work. Similar to the peer-review evaluations of papers, program committees can evaluate the reproducibility of papers that rely on empirical evidence. The results could then be published in the conference proceedings, and this would be a significant step towards open science initiatives. The availability of reports reviewing and investigating the reproducibility of previously published work would be one of the most valuable outcomes of this process. 

However, we are not there yet. Forcing such a change is not practical. Instead, it is more feasible to take small steps toward this goal. When the ideal case scenario is not feasible, the unavailability of research artifacts may cause complications for follow-up studies. In particular, if the reproduced and original results do not match with each other, it may be unclear whether this mismatch is because of missing features, bugs in the new implementation, or the irreproducibility of the original results. With this lack of clarity, debugging the real cause of any mismatched results may be quite challenging.

Hence, increasing the quantity and improving the quality of research artifacts is a good starting point for the Interspeech community. These two aspects can be addressed in parallel. We emphasize that artifacts \textbf{should} be self-contained since the previous lack of emphasis on self-containment has rendered many already-released research artifacts obsolete \cite{arvan-etal-2022-reproducibility}. Using self-contained artifacts, researchers could assess the reproducibility of previously-reported results under the same conditions. 

% This is different from the current focus on reproducibility challenges more common in the NLP and machine learning communities, in which researchers are encouraged to not use the research artifacts and instead, reproduce the results from scratch. 

Following this, finding conditions that must be controlled to facilitate effective reproducibility should be considered an important area of research. This process is also more aligned with modern software engineering principles, where testing and reviewing source code is a common practice. It also underlines the importance of good engineering practices in academia. 
% have often been ignored in the machine learning, speech and language processing communities despite the increased reliance on empirical results in modern research in these fields. 

It is understandable that researchers may not be able to share their data or research artifacts due to GDPR and confidentiality issues. These restrictions are out of the control of researchers and should be considered when evaluating the reproducibility of results. Nevertheless, sharing data sources and code should be encouraged as much as possible to achieve open science goals. Researchers who face issues about sharing their own data could perhaps consider including supplementary experiments performed on publicly available datasets (when/if possible) to strengthen their claims and foster comparison with prior work. Conference organizers could also provide support through encouraging reproducibility reports. As an example of involvement from conference leadership, ACL and EMNLP have included reproduction studies as contributions in their calls for papers. 

% \sd{We need to mention that speech data is not always available due to GDPR and confidentiality issues. However, our goal is not to fix everything in this paper but explain the current state at Interspeech in comparison to other conferences}

Conducting reproducibility evaluations may result in many failed attempts. While some of these failures may actually be due to bugs or other issues, many of them may also be due to the lack of self-containment or proper documentation. The reproducibility of a submitted research paper should be viewed as another dimension of its quality during the evaluation process. However, it is still challenging to assess the validity of research results. Similar to the peer-review evaluations, artifact evaluation is not free of flaws. A failed reproduction attempt should not single-handedly result in a paper's dismissal. However, the additional scrutiny of evaluating the potential for reproducibility of research results during the submission process would help the field progress towards open science goals in the long run. 

We summarize our recommendations for the Interspeech community below as follows:
\begin{itemize}
  \item Researchers should be encouraged to share their research artifacts.
  \item \url{isca-speech.org} could be updated to include information about the availability of research artifacts (and we are willing to share our suggestions for improvements). 
  \item Interspeech organizers could consider encouraging reproducibility evaluations for accepted papers that rely on empirical evidence.
  \item Interspeech organizers could encourage reproduction studies as a contribution or as a separate track for the annual conference.
\end{itemize}

We recognize that there may be numerous barriers to implementing any and all of these recommendations, and we encourage incremental steps toward them as well.  Nevertheless, we hope that these recommendations propel the Interspeech community forward toward producing more reproducible, open science.

\section{Limitations}

While we have done our best to ensure the accuracy of our results, there are some limitations to our work. First, we have used a simple keyword search to determine whether papers not hosted on the ACL Anthology contain links to research artifacts. This approach may have led to false positives. For example, a paper may contain a link to a GitHub repository that is not related to the research artifacts, such as a third-party dataset or tool. This may have caused an overly optimistic estimation of the percentage of papers with research artifacts published at Interspeech and NeurIPS. At the same time, since these venues have no formal procedures to include a link to the source code within a paper, some published papers may have publicly released their source code but not included any reference to that code within their paper. Unfortunately, except through mass direct inquiries, there does not seem to be a way to collect such data in a more precise manner. 

Second, we have used the \url{aclanthology.org} website to determine the availability of research artifacts for NLP conferences. We suspect that the authors of some papers may have not included the links to their research artifacts on their paper's landing page in the \url{aclanthology.org} website.  Hence, we may have underestimated the percentage of research papers with publicly available research artifacts in NLP conferences.

Finally, this work reports on the availability of research artifacts for papers published at Interspeech, which is not necessarily broadly representative of the full speech processing community.  Our focus on this conference was in part introspective, and in part a direct response to the conference theme of inclusive spoken language science and technology.  Specifically, we seek to systematically understand one aspect of research inclusivity by studying reproducibility, for which a lack thereof poses a barrier to open science.

\section{Conclusion}

In this work, we provided a high-level overview of the state of research artifact availability at Interspeech, NeurIPS, and several NLP conferences. We found that the percentage of papers with research artifacts published at Interspeech is lower than nearly all peer conferences published at selected NLP and machine learning conferences. We hope our research findings will inform the Interspeech community about the current situation around reproducibility and inspire them to take the necessary steps towards improving the quantity and quality of research artifacts in their research and at Interspeech conferences. Doing so will remove unnecessary barriers for future researchers to build upon the work presented at Interspeech and benefit the authors whose results will be reproduced.  

\section{Acknowledgements}
We thank the anonymous reviewers for their helpful feedback, which we incorporated in the final version of this manuscript.

\bibliographystyle{IEEEtran}
\bibliography{mybib,misc}

\end{document}